# Zig-zag ladders with staggered magnetic chirality in $S = 3/2$ compound $\beta$-$CaCr_2O_4$

Françoise Damay[1], Christine Martin[2], Vincent Hardy[2], Antoine Maignan[2], Gilles André[1], Kevin Knight[3], Sean R. Giblin[3], and Laurent C. Chapon[3]

[1]Laboratoire Léon Brillouin, CEA-CNRS UMR 12, 91191 GIF-SUR-YVETTE CEDEX, France

[2]Laboratoire CRISMAT, CNRS UMR 6508, 6 bvd Maréchal Juin, 14050 CAEN CEDEX, France

[3]ISIS Facility, Rutherford Appleton Laboratory-CCLRC, Chilton, Didcot, Oxfordshire OX11 0QX, United Kingdom

AUTHOR EMAIL ADDRESS francoise.damay@cea.fr

ABSTRACT

The crystal and magnetic structures of the $S = 3/2$ chain antiferromagnet $\beta$-$CaCr_2O_4$ have been investigated by means of specific heat, magnetization, muon relaxation and neutron powder diffraction between 300K and 1.5K. Owing to the original topology of the $Cr^{3+}$ magnetic lattice, which can be described as a network of triangular ladders, equivalent to chains with nearest and next-nearest neighbors interactions, evolution of the magnetic scattering intensity in this compound evidences two magnetic regimes : for 21K < T < 270K, a low-dimensionality magnetic ordering of the $Cr^{3+}$ spins is observed, simultaneously with a strong contraction of the ladder legs, parallel to *c*. Below $T_N$ = 21K, a complex antiferromagnetic ordering is evidenced, with an incommensurate propagation vector **k** = (0, 0, *q*) (*q* ~ 0.477 at 1.5K), as exchange interactions between ladders become significant. This complex magnetic ordering can be described as a honeycomb-like arrangement of cycloids, running along *c*, with staggered chiralities. The experimental observation of this staggered chirality can be understood by taking into account antisymmetric Dzyaloshinskii-Moriya exchange terms.

## INTRODUCTION

Quantum spin chains with nearest (NN, $J_1$) and next-nearest (NNN, $J_2$) neighbor antiferromagnetic interactions have attracted considerable interest to date because they exhibit a rich variety of exotic ground states depending on the $j = J_2/J_1$ ratio. Such systems are described by the model Hamiltonian :

$$\mathbf{H} = \sum_{\rho=1}^{2}\left\{ J_\rho \sum_{l} (S_l^x S_{l+\rho}^x + S_l^y S_{l+\rho}^y + \Delta S_l^z S_{l+\rho}^z) \right\} \qquad (1)$$

where $S_l$ is the spin operator at the $l$th site and $\Delta$ the exchange anisotropy.

In the classical limit $S \to \infty$, there is a crossover at $j = ¼$ between a Néel antiferromagnet and an incommensurate helical-type order with a wave-vector $k = acos[-1/(4j)]$ (equation 2). This helical order is also characterized by its chirality (defined by $\boldsymbol{\kappa}_l = \mathbf{S}_l \times \mathbf{S}_{l+1}$), that is, the sense of rotation of the spins along the helix (either left or right handed). In the quantum case, no magnetic long range order emerges for $0 \leq \Delta \leq 1$, in contrast to the classical limit. For half-odd integer $S$, it has been shown previously that there exist two distinct phases, the spin fluid and the dimer phases [1], [2]. For large $j$ values, a chiral ordered phase with gapless excitations has also been predicted in the XY case ($\Delta = 0$) using a bosonization technique [3]. The ground-state phase diagram of a $S = 3/2$ chain determined by Hikihara *et al.* using the density-matrix-renormalization group method [4] actually includes a spin-fluid to dimer ($j_d \sim 0.334$) and a dimer to gapless chiral ($j_c$) phase transition line. The dimer phase exists only for a narrow range of $j$, but the value of $j_c$ increases as $\Delta$ becomes larger, so that the dimer-chiral phase boundary tends to the Heisenberg limit for $\Delta = 1$ and $j \to \infty$, where no gapless chiral phase exists.

Amongst the few systems that realize the topology of the frustrated $J_1$-$J_2$ chain, isotypes of calcium-ferrite $CaFe_2O_4$ [5], [6] have recently been the focus of several experimental studies [7], [8], [9]. These systems are made of two-leg zig-zag ladders of transition-metal (TM) ions running along the $c$-axis, topologically equivalent to chains with competing first ($J_1$) and second neighbour ($J_2$) interactions (Figure 1a). The TM ions occupy two non-equivalent crystallographic sites, thus forming four types of zig-zag ladders, as shown on Figure 1 in the case of $\beta$-$CaCr_2O_4$ : a Cr(1) and a Cr(2) ladder and two non-equivalent Cr(1)-Cr(2) ladders ; these ladders are interconnected to form a honeycomb-like lattice in the *ab*-plane. In this series, the $S = 1$ compound $CaV_2O_4$ has been suggested as a potential candidate for the study of the chiral gapless phase, due to the observation of gapless excitations [10], [11] and, more recently, of the existence of a peak of unknown origin in the specific heat, at a temperature above the magnetic ordering transition [9].

In the present article, we report on the magnetic properties of the $S = 3/2$ compound $\beta$-$CaCr_2O_4$, studied by specific heat, magnetization, neutron diffraction and muon spin rotation on a polycrystalline sample. Even though the crystal structure of $\beta$-$CaCr_2O_4$ was first reported 70 years ago [12], to our knowledge, apart from an early study by Corliss *et al.* [13] mentioning a "complex" magnetic structure, the magnetic properties of this compound had not been investigated up to now. We show that low-dimensional magnetic correlations develop at high-temperature (T ~ 270K) and that the system becomes magnetically long-range ordered below $T_N = 21K$, although magnetic diffuse scattering is still observable. The ordered magnetic structure corresponds to a stacking of long-wavelength elliptical cycloidal modulations along the legs of the ladders, with propagation vector $\mathbf{k} = (0, 0, 0.477(1))$ and magnetic moments in the *ac*-plane. Amongst the four inequivalent ziz-zag ladders, the dominant nearest neighbour magnetic coupling term seems to connect chains of Cr ions on non-equivalent sites ($J_a$). Within ladders made of equivalent Cr sites, the cycloidal modulations along the ladder legs are of opposite chiralities, indicating the non-negligible contribution of antisymmetric Dzyaloshinskii-Moriya exchange terms.

## EXPERIMENTAL

2g of $CaCr_2O_4$ were prepared by high temperature solid state reaction. Powders of CaO and $Cr_2O_3$ were weighted in the (1:1) stoichiometric ratio and heated at 1000°C in air for a day. The powder was then pressed in the shape of pellets and heated to 1400°C for 12 hrs in air, with a subsequent annealing at 1000°C for 12 hrs in a reducing gas flow (5% $H_2$ in Ar). The samples obtained following this procedure were checked by room temperature X-ray diffraction and found to be single phase and of good crystallinity, with the expected $CaFe_2O_4$-type structure.

Magnetic susceptibility defined as $\chi$ =M/H was calculated from magnetization data measured in a magnetic field of 0.3 Tesla, on warming from 1.5 to 300K, after a zero-field cooling, using a Quantum Design SQUID magnetometer. Heat capacity measurements were carried out by means of a commercial device (PPMS, Quantum Design) using a relaxation method with a 2τ fitting procedure. Muon spin relaxation (μSR) was performed at the ISIS facility on the EMU beamline in order to further characterize any magnetic ordering on a local length scale.

Neutron powder diffraction (NPD) versus temperature was performed on the G4.1 diffractometer (λ = 2.425Å) from 1.5 to 300K, and a high resolution neutron diffractogram at 10K was recorded on the diffractometer 3T2 (λ = 1.225Å). Both diffractometers are located at the LLB (CEA-Saclay, France). High resolution neutron powder diffractograms were also recorded at 300K, 100K and 8K on the time-of-flight diffractometer HRPD (ISIS, UK). Rietveld refinements and determination of the magnetic symmetry with representation analysis were performed with programs of the FullProf suite [14].

## RESULTS AND DISCUSSION

### *I. Crystal structure and exchange topology*

The room temperature (RT) orthorhombic structure of *β*-$CaCr_2O_4$ (Figure 1), isotype with calcium ferrite $CaFe_2O_4$ [5], [6], has been refined from the HRPD data (Figure 2a) in the *Pbnm* space group with *a* = 10.6203(3)Å, *b* = 9.0801(3)Å, *c* = 2.9681(1)Å and found to be in excellent agreement with previous reports using single crystal X-ray diffraction [15]. Atomic positions and selected distances and angles are listed in Tables I and II. The $Cr^{+3}$ ions occupy two distinct crystallographic positions, both on Wyckoff position 4*c* (*x*, *y*, ¼), and are octahedrally coordinated by oxygen ions. Both octahedra are weakly distorted, with Cr-O distances ranging at 300K between 1.9805(10)Å and 2.0410(11)Å (Table II.2 and Figure 2b), the $Cr(1)O_6$ octahedron being slightly more regular ($\Delta d = 0.801\ 10^{-4}$) than the $Cr(2)O_6$ one ($\Delta d = 1.111\ 10^{-4}$). The average Cr-O distances in each $Cr^{+3}$ octahedron are only marginally different, 2.0104(8)Å for $Cr(1)O_6$ and 2.0040(8)Å for $Cr(2)O_6$. As seen on Figure 2b and Table II.2, three out of the six bonds in each $CrO_6$ octahedron are involved in close to 98° Cr(1)-O(4)-Cr(1) (or Cr(2)-O(2)-Cr(2)) angles between equivalent chromium sites, while the three other bonds connect non-equivalent Cr sites together through Cr(1)-O(1)-Cr(2) or Cr(1)-O(3)-Cr(2) bridges close to 122° and 132° respectively.

Cr(1)$O_6$ octahedra and Cr(2)$O_6$ octahedra form chains by sharing edges along the *c*-axis (Figure 1c). The crystal symmetry is such that for any given chain in the lattice (Cr(1) or Cr(2)), the Cr positions in any adjacent chain irrespective of its nature are translated along *c* by (0, 0, 1/2), thus forming zigzag ladders. Two Cr(1) in adjacent chains are in addition related by inversion symmetry. The same applies for adjacent Cr(2) chains. Cr(1) and Cr(2) chains are interconnected through shared corners, forming also two non-equivalent zig-zag ladders, depending on whether the Cr(1)-O-Cr(2) bridges are along the *a*-axis (Cr(1)-O(3)-Cr(2)) or along the *b*-axis (Cr(1)-O(1)-Cr(2)), as shown in Figure 1.

The magnetic lattice is therefore made out of four non-equivalent zig-zag ladders, interconnected to create a honeycomb-like network in the *ab*-plane (Figure 1b). The cavities of the honeycomb structure host $Ca^{2+}$ ions in prismatic coordination, slightly shifted from the centre of their coordination prisms, with Ca-O distances ranging from 2.3481(7)Å to 2.5057(9)Å (Table II.1). Figure 1 also shows the different exchange interactions terms that have to be considered, according to the crystal structure topology. Exchange interactions along the Cr chains (corresponding to second nearest-neighbor interaction $J_2$ in the ziz-zag ladder model on Figure 1a) are noted $J_{2-1}$ and $J_{2-2}$ for the Cr(1) and Cr(2) chains respectively, while first-neighbor interactions in ladders built with equivalent Cr atoms are noted $J_{1-1}$ (Cr(1)) and $J_{1-2}$ (Cr(2)), respectively. First neighbor interactions in ladders connecting Cr(1)-Cr(2) chains are noted $J_a$ and $J_b$ depending on the direction of the Cr-O-Cr bridge. The relative strengths of the magnetic exchange interactions will be discussed in the light of the determined magnetic structure but a qualitative account of the most relevant exchange paths is given in the following. Considering the short inter-atomic Cr-Cr distance of 2.9681(1)Å along the chains, magnetic interactions $J_{2-1}$ and $J_{2-2}$ are expected to be antiferromagnetic (AFM), owing to the edge-sharing octahedral configuration allowing "cation-cation" direct-exchange through the $d_{xz}$ orbitals, which are pointing along the Cr-Cr axis. Since the inter-atomic distances for Cr(1)-Cr(1) and Cr(2)-Cr(2) are identical (and equal to the *c* lattice parameter), exchange integrals $J_{2-1}$ and $J_{2-2}$ should be comparable. Therefore, whilst Cr(1) and Cr(2) sites are not related by symmetry, the ladders involving Cr(1) and Cr(2) chains can also be considered as "ideal", i.e., with a unique exchange integral ($J_2$ type) along their two legs. The exchange integrals $J_{1-1}$ and $J_{1-2}$ along the ladder rungs also involve antiferromagnetic direct-exchange, and should be smaller than $J_{2-1}$ and $J_{2-2}$, due to the larger Cr-Cr inter-atomic distances (3.0192(11)Å for Cr(1)-Cr(1) and 3.0154(11)Å for Cr(2)-Cr(2) (see Table III)). Finally the exchange interaction $J_a$ and $J_b$ between non–equivalent Cr sites are mediated by super-exchange through a Cr-O-Cr bridge ; the corresponding Cr(1)-Cr(2) $J_a$ or $J_b$ ladders are characterized by Cr(1)-Cr(2) distances about 0.4Å longer (respectively 3.6287(11)Å and 3.5615(12)Å at 300K, see Table II.2) than those involved in direct exchange. The Cr-O-Cr angle mediating the super-exchange interaction between non equivalent chromium sites varies between 132° (Cr(1)-O(3)-Cr(2) angle within the $J_a$ ladder) and 122° (Cr(1)-O(1)-Cr(2) angle within the $J_b$ ladder), which suggests a stronger coupling along the *a*-direction. A set of distances and angles summarizing the main characteristics of each of the four non-equivalent zig-zag ladders in $\beta$-$CaCr_2O_4$ is given in Table III.

The evolution of the structural parameters as a function of temperature (Figure 3), extracted from Rietveld refinement of the G4.1 neutron data, indicates that the exchange integrals are renormalized at lower temperature. There is a clear anisotropic thermal contraction, illustrated by the temperature-dependence of the relative variations of the lattice parameters with respect to the 300K values. Whilst the unit-cell volume follows a usual Debye function (inset of Figure 3), we note a small but steady increase of $\Delta a/a$ and $\Delta b/b$ and a gradual decrease of $\Delta c/c$ on cooling (Figure 3). These evolutions coincide with the appearance of short-range magnetic correlations, as will be discussed in the next section, therefore likely to be of magnetostrictive origin. In the 20K-30K temperature range, which corresponds to the occurrence of 3D magnetic ordering, *a*, *b* and *c* become almost constant, with $a$ = 10.6218(7)Å, $b$ = 9.0763(6)Å and $c$ = 2.9573(2)Å at 8K. Results of the refinement of the HRPD data at 100K and 8K (see inset of Figure 2) are summarized in Tables I, II and III. The anisotropic contraction, not coupled to any variation of the atomic positions of the chromium or oxygen atoms within the experimental error bars, has mainly an effect on the various Cr-Cr distances. It induces a definite shortening of the Cr-Cr distance along the legs of the ladders ; a simultaneous expansion of the Cr(1)-Cr(1) (or Cr(2)-Cr(2)) distances across the ladder rungs also favors higher $J_{2\text{-}1}/J_{1\text{-}1}$ and $J_{2\text{-}2}/J_{1\text{-}2}$ ratios at low temperature.

*II. Bulk magnetic properties*

Susceptibility and inverse susceptibility curves obtained from magnetization measurements between 2K and 400K are presented in Figure 4. The paramagnetic moment extracted from the Curie-Weiss dependence between 280K and 400K is 3.98(6)$\mu_B$ per Cr, slightly above the spin-only expected value of 3.87 for $S$ = 3/2 $Cr^{3+}$. The Weiss temperature $\theta_{CW}$ = -270(3)K indicates strong antiferromagnetic correlations. On cooling below 280K, the susceptibility deviates increasingly from the Curie-Weiss law, until a broad maximum is evidenced around 90K-100K, typical of low-dimensional antiferromagnetic correlations. This maximum is followed by a pronounced drop, indicating the onset of 3D antiferromagnetic correlations. In the corresponding temperature range, two peaks are actually observed on the specific heat curve $C_p(T)$ (Figure 5 and top inset of Figure 5) : the first one, at $T_{N1}$ = 21K, is associated with the onset of the 3D long range antiferromagnetic ordering in agreement with neutron diffraction, while the second one, $T_{N2}$ = 16K, is attributed to another magnetic transition, as will be discussed below. The magnetic heat capacity across both transitions, also plotted in Figure 5, has been derived by subtracting the lattice contribution estimated from fitting the high-temperature specific heat data using the first Debye approximation. The magnetic entropy (bottom inset of Figure 5) calculated using this simple approximation is close and slightly lower than the expected classical value of 2Rln4, suggesting that a fraction of the entropy is released at higher temperature, in agreement with the presence of low-dimensional correlations.

Neutron diffraction (Figure 6a) is in perfect agreement with the bulk measurements, clearly evidencing several magnetic regimes : above $T_{N1}$, a broad asymmetric feature is observed, centered around the (1, 0, ½) Bragg position (see also inset of Figure 7), which supports the picture of a low dimensionality system above the antiferromagnetic transition. It is not possible to determine the upper temperature limit of existence for this scattering, as it becomes more diffuse and comparable to the background level on warming above 180K. However, the susceptibility suggests that this effect should persist up to 280K, temperature at which the deviation from the Curie-Weiss law becomes clearly noticeable. Between $T_{N1}$ and $T_{N2}$, the diffuse scattering signal decreases and the first magnetic Bragg peaks appear, which can be indexed with a propagation vector $k = (0, 0, q)$ ($q \sim 0.479(1)$) roughly constant within this temperature range (Figure 6b). For T < $T_{N2}$, a strong decrease of $q$ is observed, along with an increase of the intensity of the magnetic Bragg peaks, which we will focus on in the next section.

$\mu$SR can be used as a local magnetic probe and, as such, is well suited for confirming the long range magnetic order and for observing its associated fluctuations. Muons are fundamental spin ½ particles, implanted randomly into the sample. The subsequent decay of the muon into a positron is dependent on the observed local field. The corresponding relaxation spectra can be fitted with an exponential of the form $A(t) = A_0 \exp(-\lambda t)$, where $A_0$ represents the amplitude of the asymmetry and $\lambda$ the fluctuation rate. In a muon relaxation experiment, long range magnetic order is evidenced by a drop in the asymmetry by 2/3, along with a slowing down of the dynamics. Figure 6c shows the temperature dependence of the normalized initial asymmetry $A_0$ and of the fluctuation rate $\lambda$. It clearly demonstrates a long range magnetic transition at ~ 21K ; a critical-type fit to the value of $A_0$ gives a magnetic ordering temperature of 21.6K. This is in agreement with the ordering at $T_{N1}$ observed using bulk techniques.

*III. Ordered magnetic structure*

*1. Representation analysis and Rietveld refinement*

Below 21K, a large fraction of the magnetic diffuse intensity is transferred into magnetic Bragg reflections, all indexed with the propagation vector **k** = (0, 0, *q*) (*q* ~ 0.477(1) at 1.5K) (Figures 6 and 7). At 1.5K, the magnetic signal is mainly found in Bragg scattering, a persistent contribution from the weak diffuse signal is, however, still observable.

The low temperature magnetic structure has been determined by Rietveld refinement using symmetry adapted modes derived from representation analysis [14]. Table IV lists the characters of the four one-dimensional irreducible representations of the little group $G_{\mathbf{k}}$. The magnetic representation $\Gamma_m$ calculated for the Wyckoff position 4*c* in the *Pbnm* space group (for both Cr sites) contains three times each representation, so that there are three basis functions for each representation. Only the symmetry modes spanned by the irreducible representation $\Gamma_3$ (following Kovalev notation [16]) provide good agreement with the experimental data. Table IV lists the basis vectors of irreducible representation $\Gamma_3$, obtained by the projection operators method. An unconstrained refinement, using only the symmetry restrictions of $\Gamma_3$, showed that the components of the magnetic moments along the crystallographic axes are nearly equivalent for both Cr sites. As a result, in the final refinement stage, these components were constrained to be equal. There are two indistinguishable moment configurations, with identical magnetic structure factors, that give a very good agreement factor $R_{mag}$ = 4.7% with the experimental data (Figure 7). They are illustrated on figure 8a. The first model (i) consists of a sinusoidal modulation of the moment amplitude, and is obtained by mixing the basis functions $\psi_1$ and $\psi_3$ (table IV) with real components. The second model (ii) describes a cycloidal elliptical modulation, obtained by mixing $\psi_1$ and $\psi_3$ with real and imaginary characters, $\psi = \psi_1 + i\psi_3$ (Table IV). For both models, the magnetic moments lay in the *ac*-plane. At 1.5K, the components of the moment are $M_x = 2.88(7)\mu_B$ and $M_z = 1.27(13)\mu_B$ along the *a* and *c* directions respectively. For model (i), this means that the wave maximum amplitude is about $3.14(8)\mu_B$, while for model (ii), the ordered component maximum is simply the major axis of the ellipse ($M_x$). Whilst it is impossible to distinguish between both models by diffraction on a polycrystalline sample, the moment value found for model (i), slightly exceeding the expected spin value of $3.0\mu_B$, seems to be non-physical. In addition, the existence of two transitions in the temperature dependence of the specific heat suggests that the cycloidal model is much more likely, and is actually reminiscent of the successive transitions seen in numerous magnetic systems presenting non-collinear complex magnetic orderings [17], [18], such as those with competing interactions in the presence of single ion anisotropy [19]. In this respect, the first transition can therefore be understood as a collinear spin-density wave with the moments oriented along the anisotropy direction (the *a*-axis, which is the major axis of the elliptical modulation), while $T_{N2}$ is indicative of the onset of a secondary order parameter along *c*, which increases the single-ion anisotropy energy but lowers the magnetic entropy dominant at low-temperature. The larger amplitude of the first peak in the heat capacity measurement, as well as the variation of the moment components with temperature ($M_x/M_z$ = 2.26 at 1.5K and $M_x/M_z$ = 2.52 at 20K) actually corroborate this assumption. A more detailed study on single crystal with closely spaced temperature points would nevertheless be necessary to confirm this model.

The magnetic point symmetry can be easily deduced from the analysis of the co-representation analysis. Since the experimental mode is a combination $\Delta_3 + i\Delta_3$, all operations of the co-little group combined with complex conjugation are lost, and the point symmetry is mm2. This symmetry can support ferroelectricity along the *c*-axis.

*2. Discussion on the cycloidal magnetic arrangement and its chirality*

Considering now that the sinusoidal model (i) is less likely to describe the magnetic structure of *β*-$CaCr_2O_4$, based on the observations stated in the previous paragraph, we will now further investigate the cycloidal magnetic arrangement [20]. The basis functions for representation $\Gamma_3$ indicate a staggered chirality of the cycloidal modulations along each leg of the ladders involving Cr atoms on equivalent sites . For example, the sign of the imaginary *z*-component of the basis function $\psi_3$ is opposite for Cr(1)-1 and Cr(1)-2, on one hand, and for Cr(1)-3 and Cr(1)-4, on the other hand (Table V). Figure 8b shows the magnetic arrangement in the *ab*-plane with symbols + and – labeling chains with opposite chiralities. Only the ladders with first-neighbor coupling $J_a$ exhibit the same sign chirality along their two legs, which is indeed the expected configuration, at least in the classical limit, considering isotropic exchange interactions [21]. It seems therefore that amongst the non-equivalent first neighbor interactions, $J_a$ is prevailing.

To take into account the staggered chirality observed within the ladders made from equivalent Cr atoms, one needs to consider a Dzyaloshinskii-Moriya (DM) interaction perturbation term, of the type $E_{DM} = D.\mathbf{S}_i \times \mathbf{S}_j$ (*i* and *j* refer to first-neighbor ions along a chain). As relativistic corrections, these terms are usually small, but can become significant when competing with a weak isotropic exchange integral. The DM interactions that are allowed in the paramagnetic group of $\beta$-$CaCr_2O_4$ are as follows : each Cr ion (Cr(1)/Cr(2)) occupies a site of ..m symmetry, and is related to an adjacent Cr ion within the same chain by a mirror operation in the *xy* plane, i.e., perpendicular to the chain axis. Since $E_{DM}$ must be invariant under the mirror plane operation, the **D** vector is directed in the *xy*-plane (Figure 9), and its *y*-component can couple to non-collinear spin arrangements in the *xz*-plane, as observed experimentally. The direction of the **D** vector is reversed by the two-fold rotation (screw axis) along the *z*-axis that relates on one hand each legs of the Cr(1)-Cr(1) ladders and each legs of Cr(2)-Cr(2) ladders on the other hand, thus explaining the reversed chirality (Figure 9). One can also easily verify that the magnetic structure determined experimentally is consistent with the transformation properties of the D vectors through the *n* and *m* mirror operations. DM interactions between Cr(1)-Cr(1) sites and Cr(2)-Cr(2) sites along the rungs are not permitted, since these atoms are related by an inversion point. For the “$J_b$ ladder”, in which both legs have also opposite chiralities, since Cr(1) and Cr(2) are not related by any symmetry operations of the paramagnetic group, the **D** vector can point in a general direction, and must also play a role in determining the opposite chiralities of this ladder. Clearly, there is a direct competition between the weak isotropic nearest-neighbor exchange interaction (such as $J_{1\text{-}1}$, $J_{1\text{-}2}$ or $J_b$), that favors uniform chirality within a ladder, and a DM interaction along the ladder leg, that favors opposite chiralities. This can be easily shown by considering a zig-zag ladder with three interactions, noted $J_1$ (first-neighbor isotropic exchange), $J_2$ (second neighbor isotropic exchange) and $D_{2y}$ (the component of the antisymmetric DM exchange pointing along *y*), and examining, in turn, a ladder with uniform chirality and the actual experimental magnetic configuration.

There are two sites, 1 and 2 (Figure 9) in the magnetic unit-cell determined experimentally, for which the magnetic moments are respectively written :

$$M_1(z) = M_x \cos(k_z z) + M_z \sin(k_z z)$$

$$M_2(z) = M_x \cos\left[k_z\left(z+\frac{1}{2}\right)\right] - M_z \sin\left[k_z\left(z+\frac{1}{2}\right)\right]$$

where *z* refers to the position along the *c*-axis, and can take any integer value for an infinite chain, $M_x$ and $M_z$ are pointing along the *x* and *z* axis respectively, and $k_z$ refers to the propagation vector component along *z* (2π x 0.477). For a modulation with uniform chirality, one simply needs to change the sign of the second term of $M_2$ in the corresponding equation. The total energy is simply obtained by summing on N cells along the *c*-direction. The umklapp terms that sum up to zero for an incommensurate modulation and infinite N are not written.

In the case of purely isotropic interactions and uniform chirality, the two energy terms that are obtained for an elliptical modulation are :

$$E_1 = NJ_1\left(M_x^2 + M_z^2\right)\cos\left(\frac{k_z}{2}\right)$$

$$E_2 = NJ_2\left(M_x^2 + M_z^2\right)\cos(k_z)$$

where $J > 0$ for antiferromagnetic interactions. One can easily find the condition on the wave vector *k* of equation 2 by derivating $[E_1+E_2]$ with respect to $k_z$.

If we now take into account the opposite chiralities observed experimentally and the DM term $D_{2y}$, the $E_2$ energy term remains unchanged, but $E_1$ is affected and there is a new term (noted $E_2'$) related to the antisymmetric component of the moments :

$$E_1 = NJ_1(M_x^2 - M_z^2)\cos\left(\frac{k_z}{2}\right)$$

$$E_2' = 2ND_{2y}(M_x M_z)\sin(k_z)$$

A ladder structure with legs having opposite chiralities is therefore stabilized only when :

$$J_1 M_z \cos\left(\frac{k_z}{2}\right) < D_{2y} M_x \sin(k_z)$$

Considering the experimental values of the magnetic parameters, this means that $D_{2y} > 0.2J_1$. The nature of the magnetic ordering in $\beta$-$CaCr_2O_4$ would therefore indicate that this condition is fulfilled for all but the "$J_a$ ladders". We propose accordingly that the magnetic properties of the lattice are mainly dictated by the corresponding $J_a$ exchange interaction, as highlighted in the bottom part of Figure 1c. The ratio of the exchange parameters $J_{2-1} = J_{2-2}$ and $J_a$ can be determined by the value of the propagation vector according to equation 2. One finds that $j = |J_{2-1}/J_a| \sim 3.3$, i.e., the system is in the large $j$ regime corresponding to the gapless chiral phase, as expected from the small incommensurability with respect to (0, 0, 1/2) [We note that a lattice translation along the *c*-axis in the *Pbnm* setting is doubled compared to the setting used for a chain with first and second neighbor interactions, so that $k = \pi q$ in equation 2]. If we now extract the isotropic exchange interactions from the Curie-Weiss law in the mean field limit, we obtain :

$$J_{2-1}+J_a = \frac{3}{2}S(S+1)\theta_{CW} = 108K \cdot$$

With $|J_{2-1}/J_a| \sim 3.3$, this leads to $J_{2-1} \sim 25K$ and $J_a \sim 83K$, which shows that long-range magnetic order sets in around $T_N \sim J_{2-1}$.

## CONCLUSION

The combination of susceptibility, heat capacity, muon relaxation and neutron scattering measurements carried out on $\beta$-$CaCr_2O_4$, which exhibits a $S = 3/2$ next and next-nearest neighbor chain magnetic topology, clearly demonstrates the existence of two magnetic regimes in this compound. As temperature decreases, an extended 1D-like regime is first observed. The second magnetic regime corresponds to the emergence of a complex Néel order, characterized by a honeycomb-like arrangement of long-wavelength cycloids running along *c*. Interestingly, the staggered chirality observed within ladders connecting symmetrically equivalent $Cr^{+3}$ sites shows that the symmetric magnetic exchange interaction within these ladder's rungs is in direct competition with antisymmetric exchange terms along the ladders legs.

## ACKNOWLEGMENTS

Authors acknowledge fruitful discussions with S. Petit (LLB). Financial support for this work was partially provided by the French Agence Nationale de la Recherche, Grant No ANR-08-BLAN-0005-01.

TABLE CAPTIONS

Table I. Structural parameters of $\beta$-$CaCr_2O_4$ at 300K, 100K and 8K (from refinements of HRPD data). The space group is *Pbnm* and all the atoms are in site 4*c* (*x*, *y*, ¼).

| | | 300K | 100K | 8K |
|---|---|---|---|---|
| | *a* (Å) | 10.6203(3) | 10.6180(6) | 10.6218(7) |
| | *b* (Å) | 9.0801(3) | 9.0737(6) | 9.0763(6) |
| | *c* (Å) | 2.9681(1) | 2.9612(2) | 2.9573(2) |
| | $V$ (Å$^3$) | 286.22(2) | 285.29(3) | 285.11(3) |
| Ca | *x* | 0.6590(2) | 0.6593(1) | 0.6593(1) |
| | *y* | 0.7598(2) | 0.7601(2) | 0.7604(2) |
| | $B$ (Å$^2$) | 0.44(3) | 0.38(3) | 0.32(3) |
| Cr(1) | *x* | 0.6126(2) | 0.6125(2) | 0.6128(2) |
| | *y* | 0.4398(2) | 0.4394(2) | 0.4397(2) |
| | $B$ (Å$^2$) | 0.29(3) | 0.23(3) | 0.17(3) |
| Cr(2) | *x* | 0.1009(2) | 0.1012(2) | 0.1008(2) |
| | *y* | 0.4165(2) | 0.4159(2) | 0.4161(2) |
| | $B$ (Å$^2$) | 0.27(2) | 0.34(3) | 0.31(5) |
| O(1) | *x* | 0.1599(1) | 0.1601(1) | 0.1603(1) |
| | *y* | 0.2026(1) | 0.2026(1) | 0.2027(1) |
| | $B$ (Å$^2$) | 0.34(2) | 0.36(2) | 0.35(2) |
| O(2) | *x* | 0.4758(1) | 0.4753(1) | 0.4755(1) |
| | *y* | 0.1165(1) | 0.1170(1) | 0.1173(1) |
| | $B$ (Å$^2$) | 0.29(2) | 0.23(2) | 0.25(2) |
| O(3) | *x* | 0.7853(1) | 0.7848(1) | 0.7849(1) |
| | *y* | 0.5267(1) | 0.5263(1) | 0.5263(1) |
| | $B$ (Å$^2$) | 0.33(2) | 0.24(2) | 0.24(2) |
| O(4) | *x* | 0.4270(1) | 0.4273(1) | 0.4272(1) |
| | *y* | 0.4180(1) | 0.4172(2) | 0.4172(2) |
| | $B$ (Å$^2$) | 0.31(2) | 0.36(2) | 0.29(3) |
| | | | | |
| | $R_{Bragg}$ (%) | 3.20 | 3.42 | 3.57 |
| | $\chi^2$ | 2.47 | 4.45 | 4.84 |

Table II. : Selected bond distances (Å) and angles (°) in $\beta$-$CaCr_2O_4$ at 300K, 100K and 8K.

*1. Ca environment*

| | 300K | 100K | 8K |
|---|---|---|---|
| Ca-O(1) (x2) | 2.4533(8) | 2.4461(14) | 2.4438(14) |
| Ca-O(2) (x2) | 2.3481(7) | 2.3402(12) | 2.3386(13) |
| Ca-O(3) | 2.4947(9) | 2.4873(16) | 2.4854(18) |
| Ca-O(3) | 2.5057(9) | 2.5052(16) | 2.5091(18) |
| Ca-O(4) (x2) | 2.3759(7) | 2.3717(14) | 2.3724(15) |

*2. Cr environment*

| | 300K | 100K | 8K |
|---|---|---|---|
| Cr(1)-O(1) (x2) | 2.0312(7) | 2.0267(13) | 2.0275(14) |
| Cr(1)-O(3) | 1.9969(10) | 1.9918(18) | 1.9801(19) |
| Cr(1)-O(4) (x2) | 2.0113(7) | 2.0161(14) | 2.0133(15) |
| Cr(1)-O(4’) | 1.9805(10) | 1.9774(18) | 1.9815(19) |
| <Cr(1)-O> | 2.0104(8) | 2.0091(15) | 2.0072(16) |
| Δd ($x10^4$) | 0.801 | 0.835 | 0.949 |
| O(1)-Cr(1)-O(4) | 177.75(5) | 177.66(8) | 177.76(8) |
| O(3)-Cr(1)-O(4’) | 162.45(5) | 162.52(8) | 162.64(9) |
| Cr(2)-O(1) | 2.0410(11) | 2.0345(22) | 2.0379(21) |
| Cr(2)-O(2) (x2) | 2.0139(7) | 2.0169(12) | 2.0128(14) |
| Cr(2)-O(2’) | 1.9906(11) | 1.9965(21) | 1.9974(21) |
| Cr(2)-O(3) (x2) | 1.9824(7) | 1.9832(12) | 1.9828(14) |
| <Cr(2)-O> | 2.0040(8) | 2.0052(15) | 2.0044(16) |
| Δd ($x10^4$) | 1.111 | 0.902 | 0.932 |
| O(2)-Cr(2)-O(3) | 172.98(5) | 172.69(9) | 172.89(9) |
| O(1)-Cr(2)-O(2’) | 173.74(5) | 173.94(9) | 174.10(9) |

$$\Delta d = \frac{1}{N}\sum_{i=1}^{N}\left(\frac{d_i - \langle d\rangle}{\langle d\rangle}\right)^2$$

Table III. Selected distances (Å) and angles (°) characterizing the four non-equivalent $Cr^{+3}$ zig-zag ladders in $\beta$-$CaCr_2O_4$ at 300K, 100K and 8K.

| | 300K | 100K | 8K |
|---|---|---|---|
| *Cr(1) ladder* | | | |
| Cr(1)-Cr(1) (*leg*) | 2.9681(1) | 2.9611(1) | 2.9573(1) |
| (*rung*) | 3.0192(11) | 3.0189(19) | 3.0209(19) |
| Cr(1)-O(4)-Cr(1) | 98.28(4) | 98.21(7) | 98.26(7) |
| Cr(1)-Cr(1)-Cr(1) | 58.88(2) | 58.74(4) | 58.61(4) |
| *Cr(2) ladder* | | | |
| Cr(2)-Cr(2) (*leg*) | 2.9681(1) | 2.9611(1) | 2.9573(1) |
| (*rung*) | 3.0154(11) | 3.0219(21) | 3.0157(22) |
| Cr(2)-O(2)-Cr(2) | 97.70(3) | 97.69(7) | 97.53(7) |
| Cr(2)-Cr(2)-Cr(2) | 58.96(3) | 58.67(5) | 58.72(5) |
| *Cr(1)-Cr(2) ladder* $J_a$ | | | |
| Cr(1)-Cr(1) (*leg*) | 2.9681(1) | 2.9611(1) | 2.9573(1) |
| Cr(1)-Cr(2) (*rung*) | 3.6287(11) | 3.6272(20) | 3.6263(21) |
| Cr(1)-O(3)-Cr(2) | 131.53(4) | 131.71(7) | 131.78(8) |
| Cr(1)-Cr(2)-Cr(1) | 48.28(2) | 48.18(3) | 48.13(4) |
| *Cr(1)-Cr(2) ladder* $J_b$ | | | |
| Cr(1)-Cr(1) (*leg*) | 2.9681(1) | 2.9611(1) | 2.9573(1) |
| Cr(1)-Cr(2) (*rung*) | 3.5615(12) | 3.5501(23) | 3.5543(23) |
| Cr(1)-O(1)-Cr(2) | 121.99(4) | 121.88(7) | 121.92(7) |
| Cr(1)-Cr(2)-Cr(1) | 49.25(2) | 49.29(4) | 49.17(4) |

Table IV. Irreducible representations of the propagation vector **k** = (0, 0, *q*) ($q \sim 0.47$) in *Pbnm*. The magnetic representation $\Gamma_m$ contains three times each irreducible representation $\Gamma_m = 3\Gamma_1 \oplus 3\Gamma_2 \oplus 3\Gamma_3 \oplus 3\Gamma_4$ ; $a = e^{+i\pi q}$ corresponds to the phase factor between $Cr^{+3}$ on each spine of a ladder.

| | 1 | 2(0, 0, ½) | *b* ¼, *y*, *z* | *n*(½, 0, ½) |
|---|---|---|---|---|
| $\Gamma_1$ | 1 | *a* | 1 | *a* |
| $\Gamma_2$ | 1 | *a* | -1 | -*a* |
| $\Gamma_3$ | 1 | -*a* | -1 | *a* |
| $\Gamma_4$ | 1 | -*a* | 1 | -*a* |

Table V. Basis functions for axial vectors associated with irreducible representation $\Gamma_3$ for Wickoff site 4*c*. $a = e^{+i\pi q}$.

| $\Gamma_3$ | (*x*, *y*, *z*) Cr(1)-1 | (-*x*+1, -*y*+1, *z*+½) Cr(1)-2 | (-*x*+³⁄₂, *y*+½, *z*) Cr(1)-3 | (*x*-½, -*y*+½, *z*+½) Cr(1)-4 |
|---|---|---|---|---|
| $\psi_1$ | (1 0 0) | (a* 0 0) | (-1 0 0) | (-a* 0 0) |
| $\psi_2$ | (0 1 0) | (0 a* 0) | (0 10 ) | (0 a* 0) |
| $\psi_3$ | (0 0 1) | (0 0 -a*) | (0 0 1) | (0 0 -a*) |

FIGURE CAPTIONS

Figure 1 (color online) : (a) Schematic diagram of the zig-zag triangular ladder, the theoretical equivalent chain is also reported in the lower part to show the first ($J_1$) and second nearest-neighbor ($J_2$) exchange interactions (b) Crystal structure of $\beta$-$CaCr_2O_4$ projected along [001], and (c) along [100] and [010]. The two distinct $Cr^{3+}$ crystallographic sites Cr(1) and Cr(2), octahedrally coordinated by oxygen atoms, are shown as green and blue polyhedra, respectively. The different magnetic exchanges paths are shown by arrows (see text for details). The red bonds show the ladders rungs between adjacent chains on non-equivalent sites (Cr(1)/Cr(2)).

Figure 2 : (a) Rietveld refinement of time of flight neutron diffraction data recorded on HRPD (ISIS) of $\beta$-$CaCr_2O_4$ at 300K and (inset) 8K (experimental data : open circles, calculated profile : continuous line, allowed Bragg reflections : vertical marks. The difference between the experimental and calculated profiles is displayed at the bottom of each graph). (b) Geometries (with corresponding distances at 300K) of the two types of $CrO_6$ octahedra encountered in $\beta$-$CaCr_2O_4$, $Cr(1)O_6$ and $Cr(2)O_6$.

Figure 3 : Evolution with temperature of the $\beta$-$CaCr_2O_4$ normalised cell parameters *a*, *b* and *c*. Inset : Temperature evolution of the corresponding cell volume *V*.

Figure 4 : Temperature evolution of the susceptibility and inverse susceptibility of $\beta$-$CaCr_2O_4$ in 0.3T. The shaded area between the paramagnetic regime (P) and the long-range magnetic ordering (3D) corresponds to the onset of low dimensionality ordering as observed by neutron diffraction (G4.1 NPD data).

Figure 5 : Specific heat $C_p$ versus temperature of $\beta$-$CaCr_2O_4$. The open circles and dotted line represent respectively the experimental data points and the lattice Debye contribution fitted from high-temperature data. Top inset : enlargement in the vicinity of $T_N$ of $C_p(T)$. Bottom inset : temperature variation of the magnetic entropy. The expected classical value per formula unit is shown as a horizontal line.

Figure 6 (color online) : (a) Evolution with temperature of the neutron diffractograms (G4.1) of *β*-$CaCr_2O_4$. $T_{N1}$ and $T_{N2}$ refer to the transition temperatures observed on the specific heat data. (b) Temperature dependence of the *q* component of the magnetic propagation vector *k* = (0, 0, *q*). (c) Temperature dependence of the normalized initial asymmetry (red circles) and fluctuation rate (λ, black solid squares) obtained from μSR. Long range magnetic order is observed using this local probe at around 21K, in agreement with the bulk susceptibility. The solid line is a fit using a critical exponent to obtain an approximate ordering temperature.

Figure 7 : Low angle enlargement of the Rietveld refinement of the neutron diffraction data (G4.1) of *β*-$CaCr_2O_4$ at 1.5K (experimental data : open circles, calculated profile : continuous line, allowed Bragg reflections : vertical marks. The difference between the experimental and calculated profiles is displayed at the bottom of each graph). Inset : enlargement of the low dimensionality magnetic scattering feature at 30K and 100K.

Figure 8 : (a) Sinusoidal (*i*) and cycloidal (*ii*) magnetic spin configurations along equivalent chromium ladders. (b) (upper part) Distribution of the sign of the chirality along *c* within Cr chains in the honeycomb lattice. Two adjacent chains with same chirality are outlined in grey, the corresponding Cr(1)-Cr(2) ladder $J_a$ is also illustrated (lower part).

Figure 9 : Schematic representation of a zig-zag ladder between equivalent Cr atoms with its symmetry elements, and of its magnetic cycloidal modulation (shown as elliptical envelops) in the *ac*-plane ($M_x$ and $M_z$). Chiralities are opposite between the two chains (atoms 1 and atoms 2). The isotropic exchange nearest and next-nearest interactions $J_1$ and $J_2$, as well as the antisymmetric DM vectors ($D_2$), are shown.

Figure 1

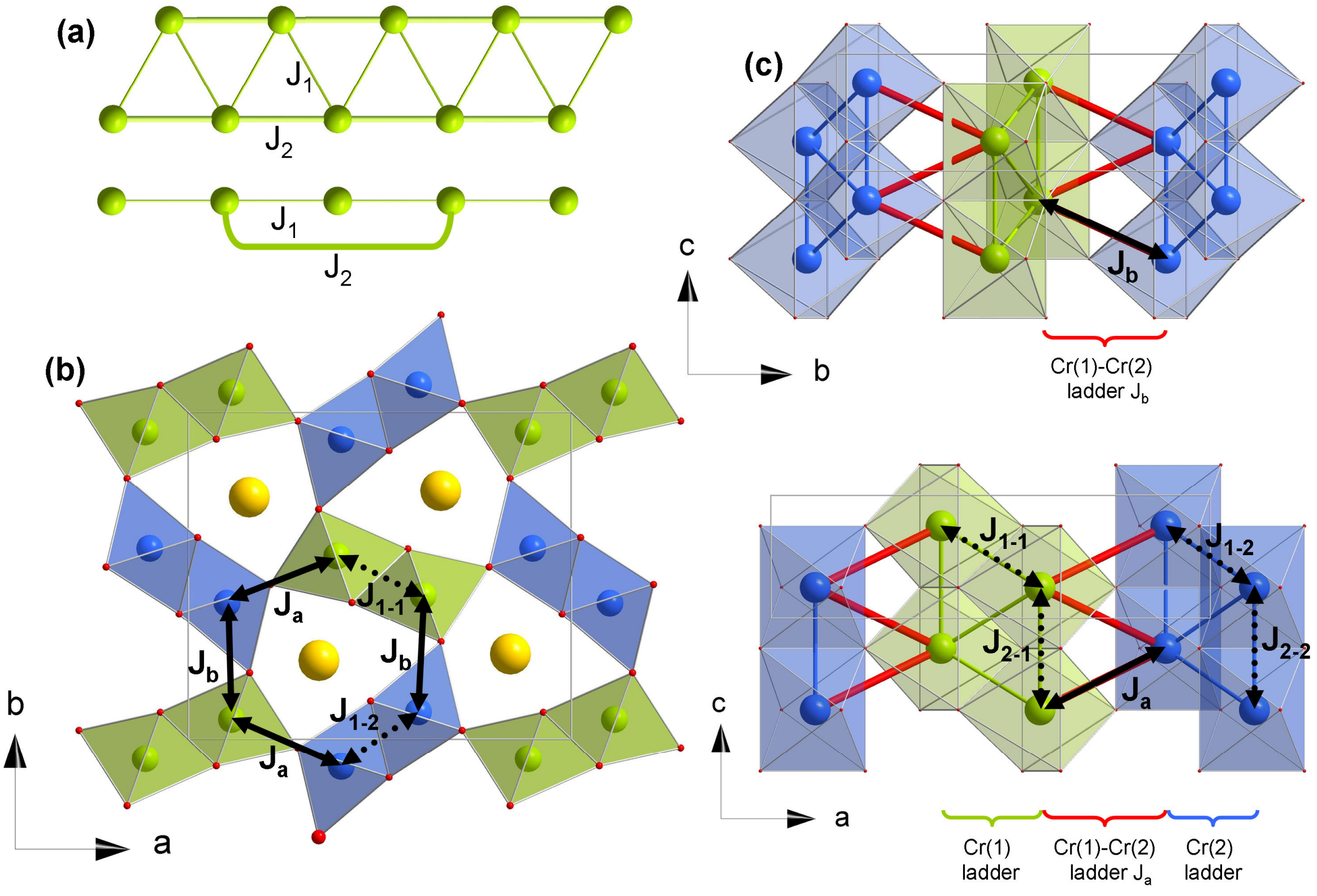

Figure 2a

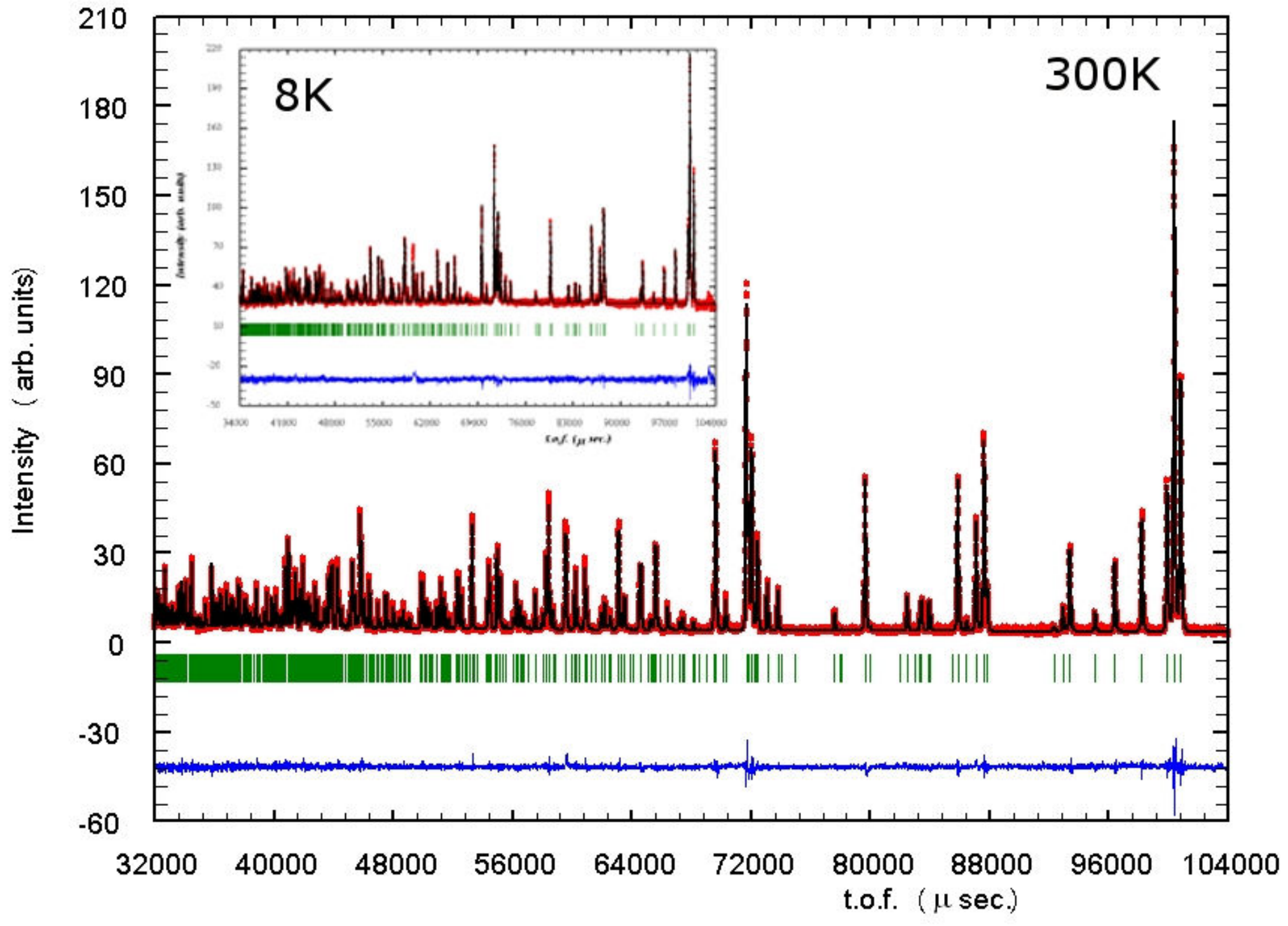

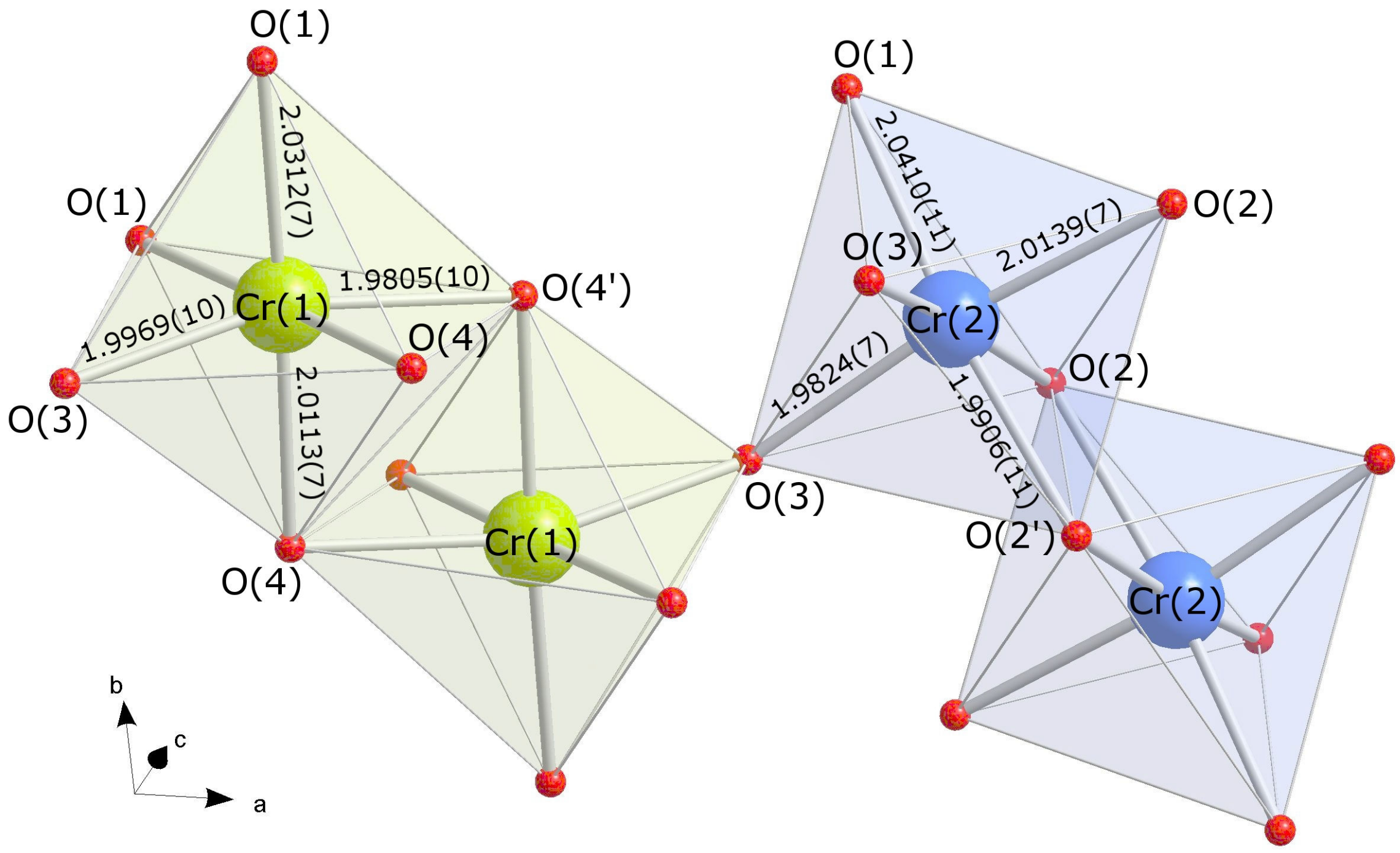

Figure 2b

Figure 3

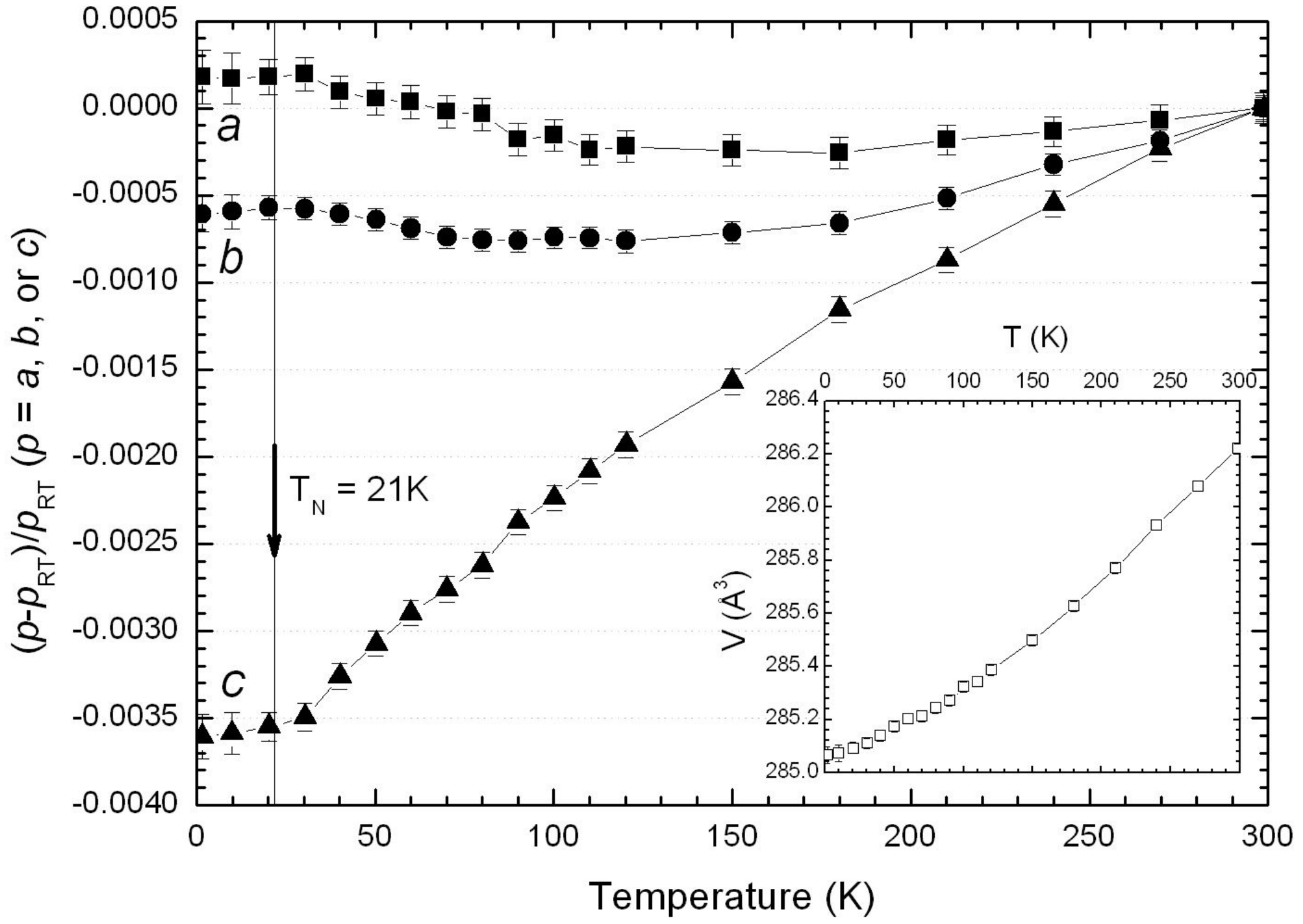

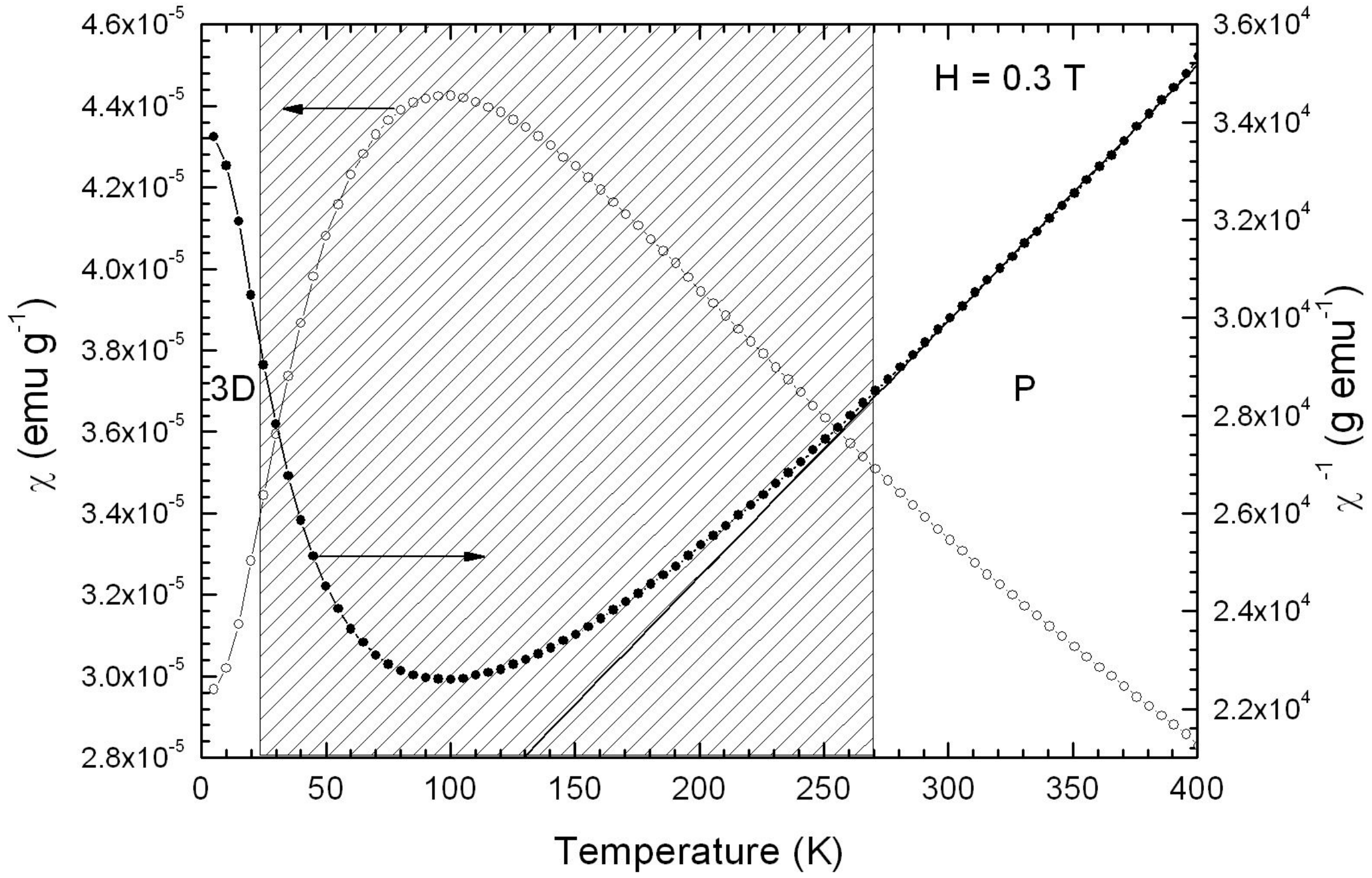

Figure 4

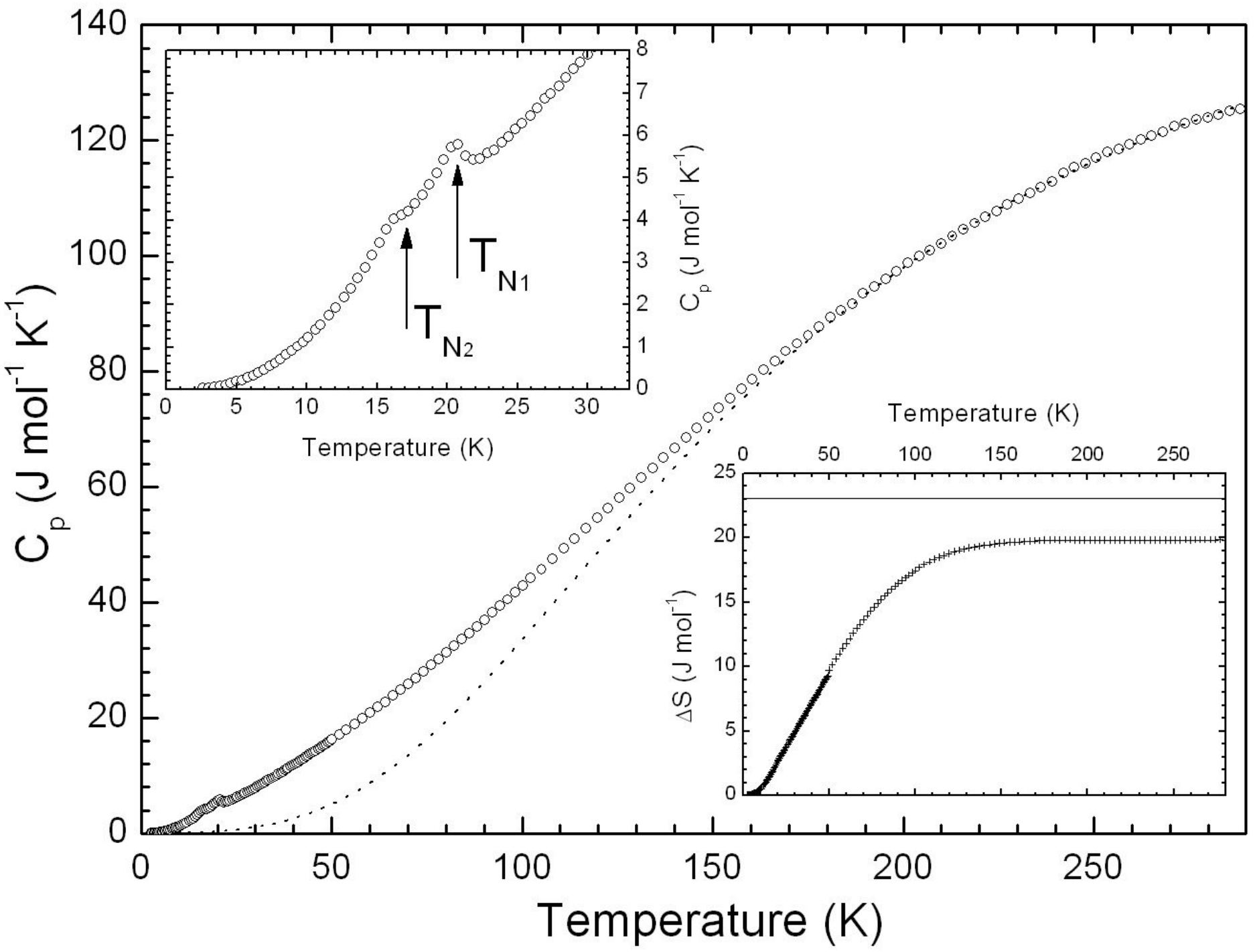

Figure 5

Figure 6

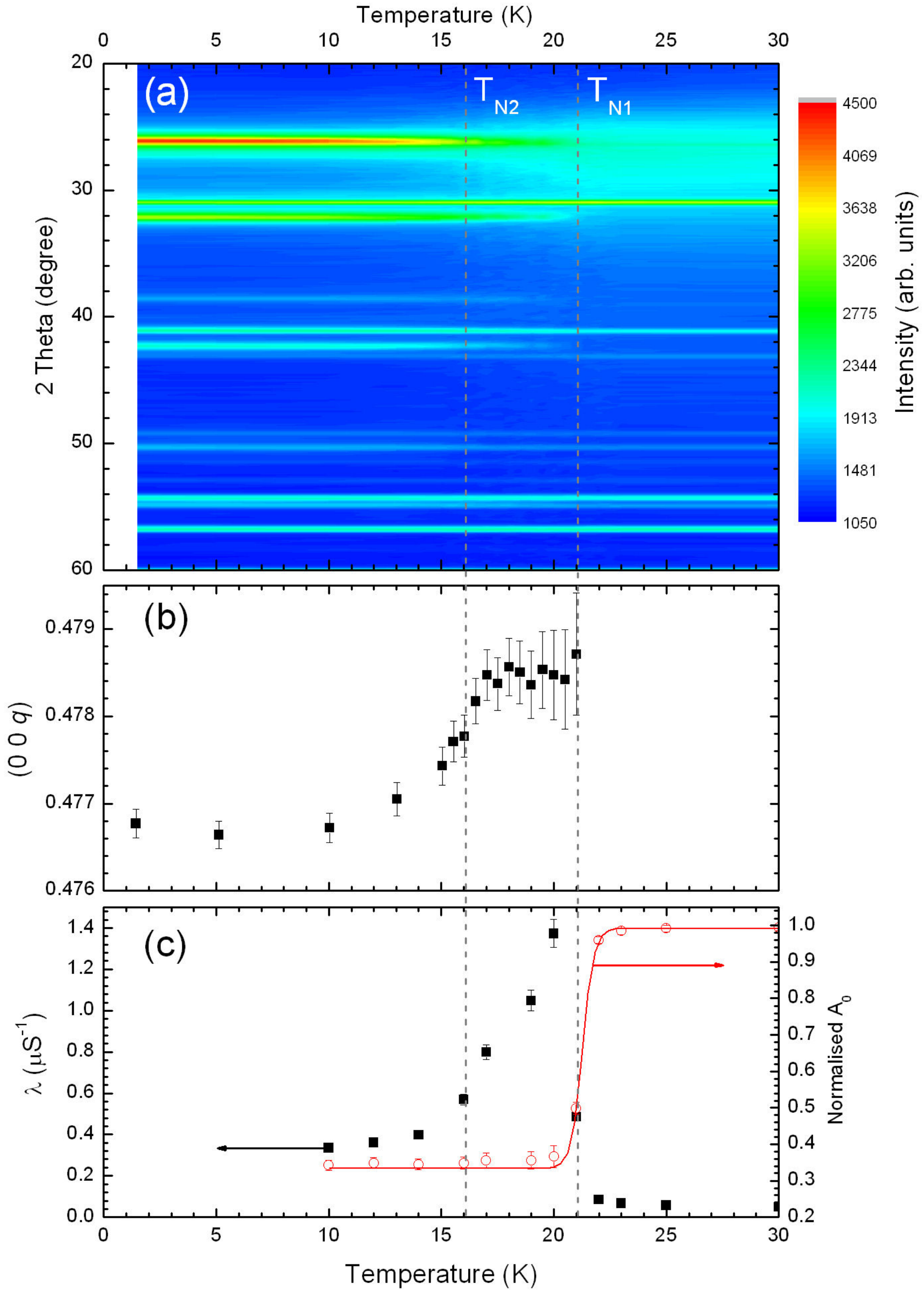

Figure 7

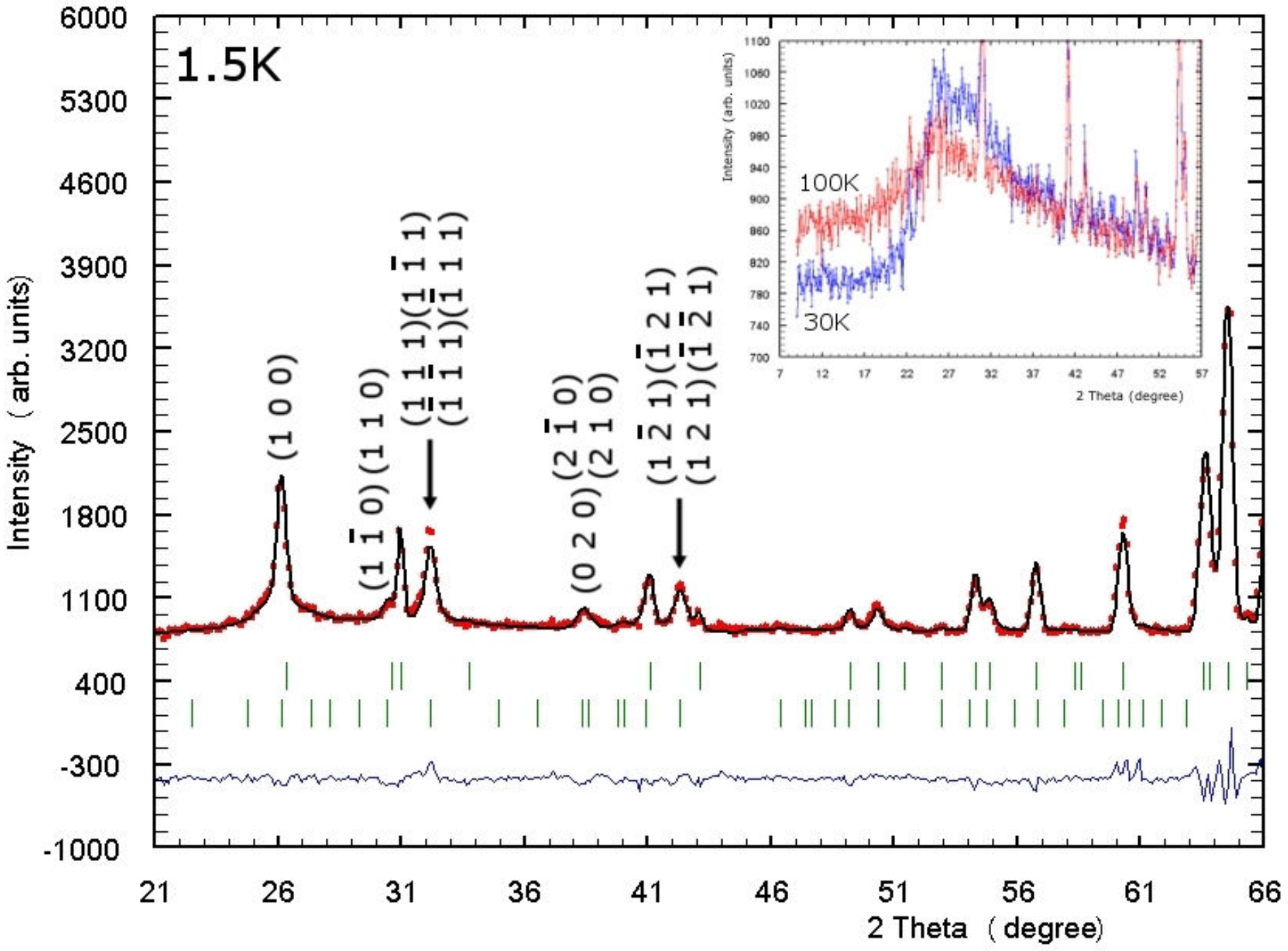

Figure 8

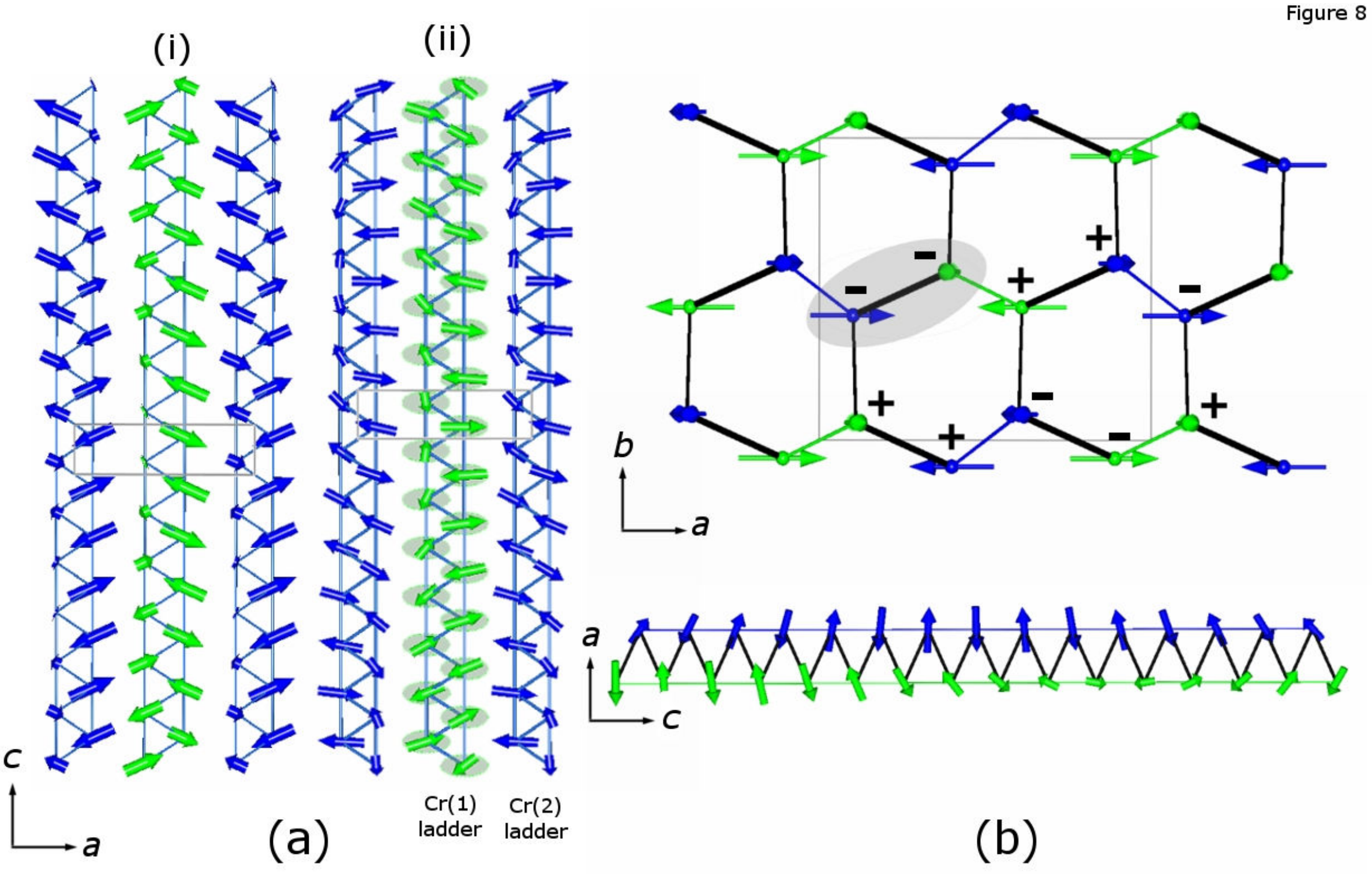

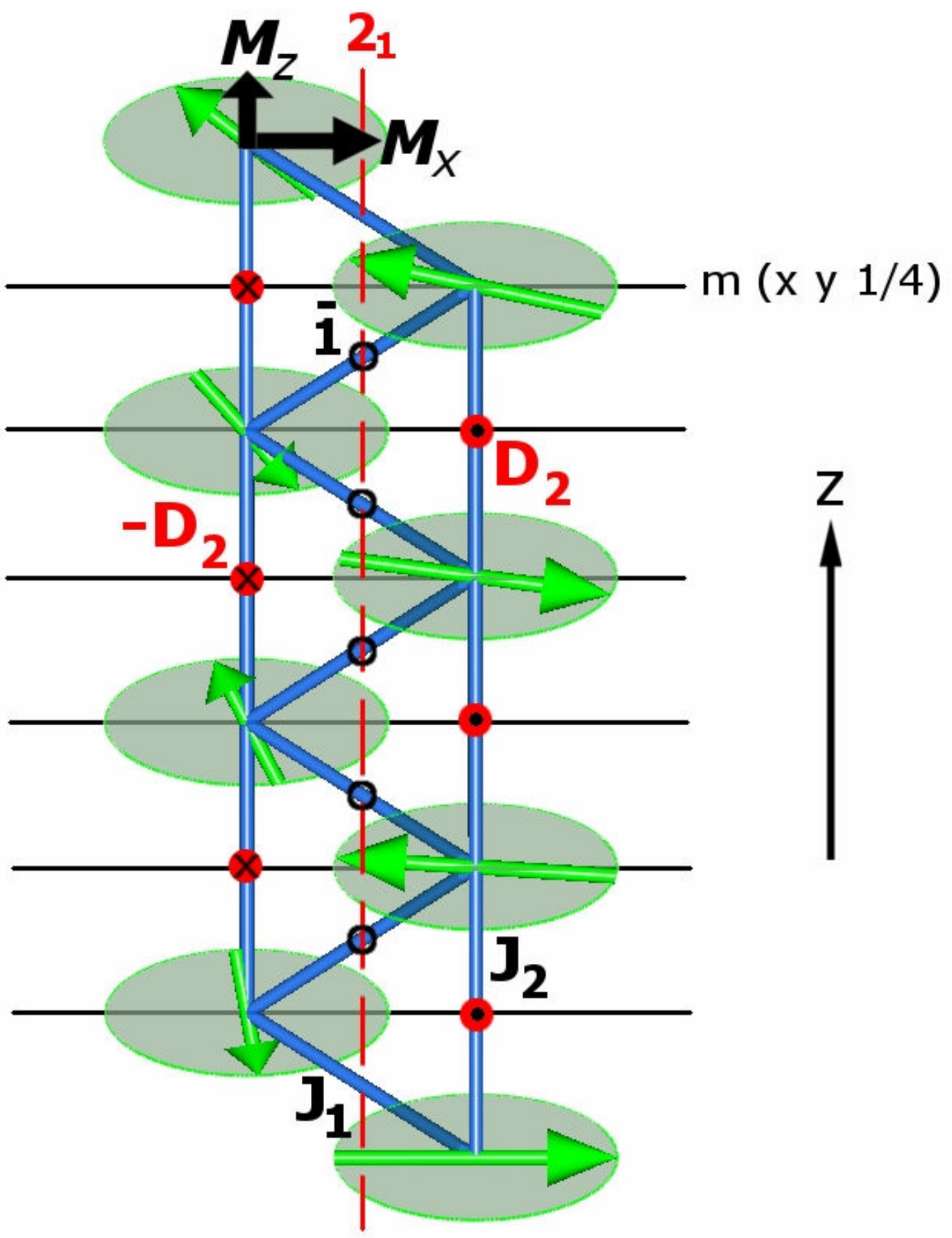

**Figure 9**